**Fast, accurate, and transferable many-body interatomic potentials by symbolic regression**

Alberto Hernandez[1], Adarsh Balasubramanian[1], Fenglin Yuan[1], Simon Mason[1], and Tim Mueller[1, *]

[1] Department of Materials Science and Engineering, Johns Hopkins University, Baltimore, 21218 MD, USA

Correspondence: Tim Mueller (tmueller@jhu.edu)

**ABSTRACT**

The length and time scales of atomistic simulations are limited by the computational cost of the methods used to predict material properties. In recent years there has been great progress in the use of machine learning algorithms to develop fast and accurate interatomic potential models, but it remains a challenge to develop models that generalize well and are fast enough to be used at extreme time and length scales. To address this challenge, we have developed a machine learning algorithm based on symbolic regression in the form of genetic programming that is capable of discovering accurate, computationally efficient many-body potential models. The key to our approach is to explore a hypothesis space of models based on fundamental physical principles and select models within this hypothesis space based on their accuracy, speed, and simplicity. The focus on simplicity reduces the risk of overfitting the training data and increases the chances of discovering a model that generalizes well. Our algorithm was validated by rediscovering an exact Lennard-Jones potential and a Sutton Chen embedded atom method potential from training data generated using these models. By using training data generated from density functional theory calculations, we found potential models for elemental copper that are simple, as fast as embedded atom models, and capable of accurately predicting properties outside of their training set. Our approach requires relatively small sets of training data, making it possible to generate training data using highly accurate methods at a reasonable computational cost. We present our approach, the forms of the discovered models, and assessments of their transferability, accuracy and speed.

**INTRODUCTION**

In recent years there have been great advances in the use of machine learning to develop interatomic potential models.[1-20] In this approach, the development of an interatomic potential model is treated as a supervised learning problem,[21] in which an optimization algorithm is used to search a hypothesis space of possible functions to find those that best reproduce the energies, forces, and possibly other properties of a set of training data. Potential models developed in this way are often able to achieve accuracy close to that of the method used to generate the training data, with linear scalability and orders of magnitude increase in performance. Alternatively, potential models may be generated by using fundamental physical relationships to derive a simple parameterized function. The parameters of this function are typically then fit to a smaller set of training data. Examples of potential models generated using this latter approach include the embedded atom method (EAM) and bond-order potentials.[22-28]

There are advantages and disadvantages to both approaches to potential model development. Machine learning can be used to develop models for a wide variety of different chemical systems, and because many machine learning algorithms explore a large hypothesis space, they are often able to achieve very high levels of accuracy on structures where the local environments of the atoms are similar to those that are contained in the data used to train the model.[1-3] On the other hand, models developed from fundamental physical relationships are often simpler and orders of magnitude faster than machine learning potential models,[29] allowing them to be used to model systems at much longer time and length scales. Because they are derived from physics, they can be expected to perform relatively well when they encounter local environments that are unlike the ones they were trained on. The hypothesis space of these potential models is relatively small compared to most machine learning potentials, meaning that less data is required to train them but also that they are typically unable to achieve the same level of accuracy as many potentials generated using machine learning.

Here we present a hybrid approach in which machine learning is used to develop simple, fast potential models. Previous work has demonstrated that genetic algorithms can be effectively used to find parameters for interatomic potential models with known functional forms.[30-35] Our approach adds to these efforts by identifying new functional forms for the models themselves. To accomplish this we use symbolic regression as implemented using genetic programming, in which simple expressions for the potential energy surface are optimized by simulating the process of natural selection.[36,37] Genetic programming has been used to rediscover fundamental physical laws[38] and applied in materials science to find descriptors of complex material properties.[39,40] It has also previously been used to identify simple interatomic potentials.[9,41-43] Here we go beyond these previous efforts by demonstrating that genetic programming is capable of finding fast, accurate and transferable many-body potentials for a metallic system from *ab-initio* calculations.

The key to our approach is the construction of a physically meaningful hypothesis space, achieved by analyzing interatomic potentials that were derived from physical principles.[25,27,28] We take advantage of natural similarities in the functional forms of simple, physics-derived models[25] to construct a hypothesis space that contains many such functional forms. The hypothesis space that we use consists of all functions that can be constructed from combinations of addition, subtraction, multiplication, division, and power operators; constant values and distances between atoms; and an operator that performs a sum over functions of distances between a given atom and all neighbors within a given cutoff radius. This space contains a wide variety of potential models derived from fundamental physical interactions, including nearly all pair potentials (e.g. Lennard-Jones,[44] Coulomb,[45] Morse[46]) as well as many-body glue potentials,[25] bond-order potentials (without the bond angle terms),[25,26,47,48] and combinations thereof. Even for relatively simple hypothesis spaces such as this one, it is difficult to enumerate a list of even relatively simple functional forms that can be created due to the large number of ways in which the various operators and values can be combined.[40] Here we use a genetic algorithm and multi-objective optimization to search this hypothesis space for interatomic potentials that are simple (and thus more likely to be generalizable[49]), fast, and accurate. Additional details of our approach are provided in the Methods section.

**RESULTS**

*Validating the machine learning algorithm*

To validate our algorithm, we tested its ability to rediscover the exact form of two interatomic potentials: the Lennard-Jones potential and the Sutton-Chen (SC) EAM potential. In each case, the genetic algorithm was able to identify the exact function used to generate the training data. The training data for the Lennard-Jones potential were generated by taking 75 snapshots (1 snapshot every 5000 steps with a time step of 1 fs) of 32-atom molecular dynamics simulations: 15 snapshots at 80 K (NVT), 15 snapshots at 80 K and 100 kPa (NPT), 15 snapshots at 100 K (NVT), 15 snapshots at 100 K and 100 kPa (NPT) and 15 snapshots at 20,000 K (NVT). It consisted of 75 energies and 7200 components of force[50], generated using the following parameterized model for argon[51]:

$$V_{LJ} = \sum_i \sum_j \left( \frac{49304.15}{r^{12}} - \frac{34.88}{r^6} \right) \tag{1}$$

where $V_{LJ}$ is the potential energy of the system, the index i represents an atom in the structure, j is its neighbor and r is the distance between the two atoms. The genetic programming algorithm found:

$$V = \sum_i \left( -50.18(983.04) \left( \sum_j (3.35r)^{-6.00} - \sum_j r^{-12.00} \right) \right) \quad (2)$$

which simplifies to the form of the Lennard-Jones potential in equation (1).

The training data for the SC EAM potential were obtained from 100 snapshots (1 snapshot every 100 steps with a time step of 1 fs) of 32-atom molecular dynamics simulations: 25 snapshots at 300 K, 25 snapshots at 1600 K, 25 snapshots at 3800 K and 25 snapshots at 20,000 K, all in the NVT ensemble. The training set consisted of 100 energies and 9600 components of force. The potential used to generate the training data was parametrized for copper:

$$V_{SC} = \sum_i \left( \sum_j \frac{644.52}{r^9} - \left( \sum_j \frac{527.62}{r^6} \right)^{0.5} \right) \quad (3)$$

The artificial intelligence algorithm found:

$$V = \sum_i \left( -0.73 - 2.53 \left( \left( -0.66(384.39) \sum_j r^{-9.00} \right) + \left( 0.25 / \left( 20.63 \sum_j r^{-6.00} \right) \right)^{-0.50} \right) \right) \quad (4)$$

When it is simplified, it gives the same form as $V_{SC}$ with a constant shift and a slight difference between the constant parameters that could be eliminated by tightening the convergence criterion for parameter optimization. The values of the parameters in the exponents were found to the second decimal place.

*Discovering new models for copper*

Having established that our genetic programming algorithm can find the exact form of simple pair and many-body potentials, we evaluated its ability to find potential models from data generated using density functional theory[52] (DFT). For this purpose, we generated 150 snapshots (1 snapshot every 100 steps with a time step of 1 fs) of 32-atom DFT molecular dynamics simulations on fcc copper: 50 snapshots at 300K (NVT), 50 snapshots at 1400 K (NVT) and 50 snapshots at 1400 K (NPT at 100 kPa). The copper had melted and lost its fcc structure for the simulations at 1400 K. The data consisted of 150 energies, 14400 components of forces and 900 components of virial stress tensors[53]. One half was randomly selected for training and the other half for validation. Models were evaluated on three metrics: complexity, defined as the number of nodes on the model; computational cost, defined as the number of summations over neighbors, as these typically consume most of the execution time; and fitness, defined as a weighted sum of the mean squared errors of the energies, forces and stresses, which were normalized to unitless values as described in the methods section:

$$fitness = 1000 * (0.5 MSE_{energy} + 0.4 MSE_{force} + 0.1 MSE_{stress}) \quad (5)$$

To identify promising models we constructed a three-parameter convex hull based on fitness, computational cost, and complexity. Some of the models on this hull are shown in Table 1.

Table 1. The 3-dimensional convex hull of models found by the machine learning algorithm

| Fitness | Cost* | Complexity | Expression |
|---|---|---|---|
| 5393157 | 1 | 2 | $\sum rf(r)$ |
| 1800.1 | 1 | 4 | $\sum r^{-3.20} f(r)$ |
| 105.30 | 1 | 8 | $\sum (649.17 r^{-9.83} - 0.09) f(r)$ |
| 54.144 | 1 | 10 | $\sum (r^{10.20-5.49r} - 0.07) f(r)$ |
| 26.906 | 2 | 13 | $\sum r^{10.20-5.49r} f(r) + 33.77 \left( \sum f(r) \right)^{-1}$ |
| 8.1584 | 2 | 15 | $\sum r^{10.21-5.48r} f(r) + 1.19 \left( \sum 0.33^r f(r) \right)^{-1}$ |
| 7.8230 | 2 | 21 | $\sum (r^{10.21-5.47r} - 0.21^r) f(r) + 0.97 \left( \sum 0.33^r f(r) \right)^{-1}$ |
| 7.8229 | 2 | 25 | $0.999 \sum (r^{10.21-5.46r} - 0.21^r) f(r) + 0.97 \left( \sum 0.33^r f(r) \right)^{-1} + 5.76$ |
| 7.4131 | 4 | 19 | $\sum r^{10.21-5.48r} f(r) + \left( 3.07 \sum f(r) \right) \left( \sum 0.31^r f(r) \right)^{-1} \left( \sum rf(r) \right)^{-1}$ |
| 4.7294 | 3 | 28 | $7.33 \sum r^{3.98-3.94r} f(r) + \left( 27.32 - \sum (11.13 + 0.03 r^{11.74-2.93r}) f(r) \right) \left( \sum f(r) \right)^{-1}$ |
| 4.2932 | 4 | 29 | $6.76 \sum r^{4.00-3.88r} f(r) + 17.25 \left( \sum f(r) \right) \left( \sum r^{11.68-3.07r} f(r) \right)^{-1} + 25.30 \left( \sum f(r) \right)^{-1}$ |

Notes: the models with fitness 7.8230 and 4.7294 are named GP1 and GP2 respectively. "Cost" is based on the number of summations. $f(r)$ is the smoothing function defined in equation (7).

Table 2. Interatomic potentials near the Pareto frontiers in Figure 3.

| Name | Expression |
|---|---|
| SC[54] | $E_i = \sum_j \dfrac{644.52}{r^9} f(r) - \left( \sum_j \dfrac{527.62}{r^6} f(r) \right)^{0.5}$ |
| GP1 | $E_i = \sum_j (r^{10.21-5.47r} - 0.21^r) f(r) + 0.97 \left( \sum_j 0.33^r f(r) \right)^{-1}$ |
| GP2 | $E_i = 7.33 \sum_j r^{3.98-3.94r} f(r) + \left( 27.32 - \sum_j (11.13 + 0.03 r^{11.74-2.93r}) f(r) \right) \left( \sum_j f(r) \right)^{-1}$ |
| EAM2[55] | $E_i = \sum_j E_1 \left( e^{-2\alpha(r-r_0)} - 2e^{-\alpha(r-r_0)} \right) f(r) + F\left( \sum_j r^6 (e^{-\beta r} + 2^9 e^{-2\beta r}) f(r) \right)$ <br><br> $\sum_i F(\bar{\rho}_i) = E(L) - \dfrac{1}{2} \sum_i \sum_j E_1 \left( e^{-2\alpha(r-r_0)} - 2e^{-\alpha(r-r_0)} \right) f(r)$ <br><br> $E(L) = -E_{sub}(1+a^*) e^{-a^*}$ <br><br> $a^* = (a/a_0 - 1)/(E_{sub}/9B\Omega)^{1/2}$ |

| ABCHM[56] | $E_i = \sum_j \varphi(r)f(r) + 1.57 \cdot 10^{-5} \left(\sum_j \psi(r)f(r)\right)^2 - \left(\sum_j \psi(r)f(r)\right)^{0.5}$ |
|---|---|
| | $\varphi(r) = \begin{cases} e^{0.82+16.01r-15.73r^2+3.80r^3}, & 1 < r < 1.9 \\ +0.62(4.43-r)^3, & 1.9 < r < 4.43 \\ -3.02(4.17-r)^3, & 1.9 < r < 4.17 \\ +2.84(4.04-r)^3, & 1.9 < r < 4.04 \\ -0.41(3.62-r)^3, & 1.9 < r < 3.62 \\ +0.65(3.13-r)^3, & 1.9 < r < 3.13 \\ +0.81(2.56-r)^3, & 1.9 < r < 2.56 \end{cases}$ $\psi(r) = \begin{cases} 0.21(4.43-r)^3, & 1.9 < r < 4.43 \\ +0.36(3.62-r)^3 & 1.9 < r < 3.62 \end{cases}$ |
| CuNi[57] | $E_i = \frac{1}{2}\sum_j \left(D_M \left[1-e^{-\alpha_M(r-R_M)}\right]^2 - D_M\right)f(r) + F(\bar{\rho}_i)$ $\bar{\rho}_i = \sum_j \tanh(20r^2)\left\{r^6\left(e^{-\beta r} + 2^9 e^{-2\beta r}\right) + \frac{\sigma^{(1)}}{\mu^{(1)}}e^{-\frac{1}{2}\left[\mu^{(1)}(r-R_B)\right]^2} - 0.1\sigma^{(1)}e^{-\frac{1}{2}\left[\mu^{(1)}(r-(R_B+0.5))\right]^2}\right\}f(r)$ $\sum_i F(\bar{\rho}_i) = E(L) - \frac{1}{2}\sum_i\sum_j \left(D_M\left[1-e^{-\alpha_M(r-R_M)}\right]^2 - D_M\right)f(r)$ $E(L) = -E_{sub}(1+a^*)e^{-a^*}$ $a^* = (a/a_0 - 1)/(E_{sub}/9B\Omega)^{1/2}$ |
| EAM1[55] | $E_i = \sum_j \left(\begin{bmatrix} E_1\left(e^{-2\alpha_1(r-r_0^{(1)})} - 2e^{-\alpha_1(r-r_0^{(1)})}\right) + \\ E_2\left(e^{-2\alpha_2(r-r_0^{(2)})} - 2e^{-\alpha_2(r-r_0^{(2)})}\right) + \delta \end{bmatrix} f(r) - \sum_{n=1}^{3}\left(H\left(r_s^{(n)}-r\right)S_n(r_s^{(n)}-r)^4\right)\right) + F(\bar{\rho})$ $if(\bar{\rho}_i < 1): F(\bar{\rho}_i) = F^{(0)} + 0.5F^{(2)}(\bar{\rho}_i-1)^2 + \sum_{n=1}^{4}\left(q_n(\bar{\rho}_i-1)^{n+2}\right)$ $else: F(\bar{\rho}_i) = \frac{F^{(0)} + 0.5F^{(2)}(\bar{\rho}_i-1)^2 + q_1(\bar{\rho}_i-1)^3 + Q_1(\bar{\rho}_i-1)^4}{1+Q_2(\bar{\rho}_i-1)^3}$ $where: \bar{\rho}_i = \sum_j\left(\left[ae^{-\beta_1(r-r_0^{(3)})^2} + e^{-\beta_2(r-r_0^{(4)})}\right]f(r)\right)$ |

Note: All potentials are in units of eV and Å. $f(r)$ is a smoothing function; for GP1 and GP1 it is defined in equation (7). EAM2 and CuNi defined the embedding function to match a universal equation of state[58].

Many of the models discovered by the genetic programming algorithm have forms that resemble the embedded atom model, or "glue" type potentials. The models consist of a sum of a pairwise term with a repulsive component and a many-body "glue" type attractive term which consists of a nonlinear transformation (an "embedding" function) of a sum over neighbors (the "density"). Here we select two of the models, which we label GP1 and GP2, for further analysis based on their favorable tradeoff between simplicity and their prediction errors for the elastic constants (Table S9). In GP1, the simpler of the two models, the embedding function is simply the inverse of the density. In GP2, the embedding function is the same, and it is multiplied by a sum of pairwise interactions to form the glue term. Although GP1 and GP2 resemble known potential models, there are some notable differences. They have much simpler functional forms than most other copper potential models, and they have a different form for the attractive "glue" part of the potential. It is common in EAM-type potential models for the embedding function to be the negative square root of the density; this can be derived from the second moment approximation.[25] In GP1 and GP2, the attractive term instead depends on the positive inverse of a sum over pairwise interactions. Unlike the other models, this embedding function is bounded in the limit of high densities and diverges to infinity in the limit of zero density. GP1 and GP2 also include terms with the unusual form of $r^{a-br}$, which grows by a power law before decaying superexponentially. The resulting models demonstrate high predictive power for condensed phases and defects that were not included in the training data and, even though there were no surfaces in the data used to train them, they largely avoid the severe underprediction of surface energies that are common for embedded-atom type models (Table S6).[55]

*Validating and evaluating the transferability of the interatomic potentials*

As might be expected by their simplicity, neither GP1 nor GP2 overfit their training data. For each model, there is little difference between the training mean absolute error and validation mean absolute error for energies, components of force vectors and components of the virial stress tensors (Figure 1). Both models similarly reproduce the radial distribution function of a liquid state well (Figure 2), which is likely partially due to the inclusion of snapshots of the liquid state in their training data. As an initial comparison of the performance between GP1, GP2, and other similar potential models, we evaluate how well they predict the elastic constants of fcc copper. The elastic constants C11, C22, and C44 are a widely used benchmark of copper potential model performance, allowing us to make a comparison between nine different copper potential models for which elastic constant data is available. We have plotted the maximum percent error in predicted elastic constants against the complexity of the model, as measured by number of nodes, in Figure 3. These errors, and all errors listed in this paper, are measured against each model's own target values, which are provided in the Supplementary Information. The potentials discovered by the machine learning algorithm presented in this work significantly change the Pareto frontier of interatomic potentials, defined as the set of interatomic potentials for which no other potential has less error and is less complex. They have errors comparable to the most accurate potential models and complexity comparable to the simplest.

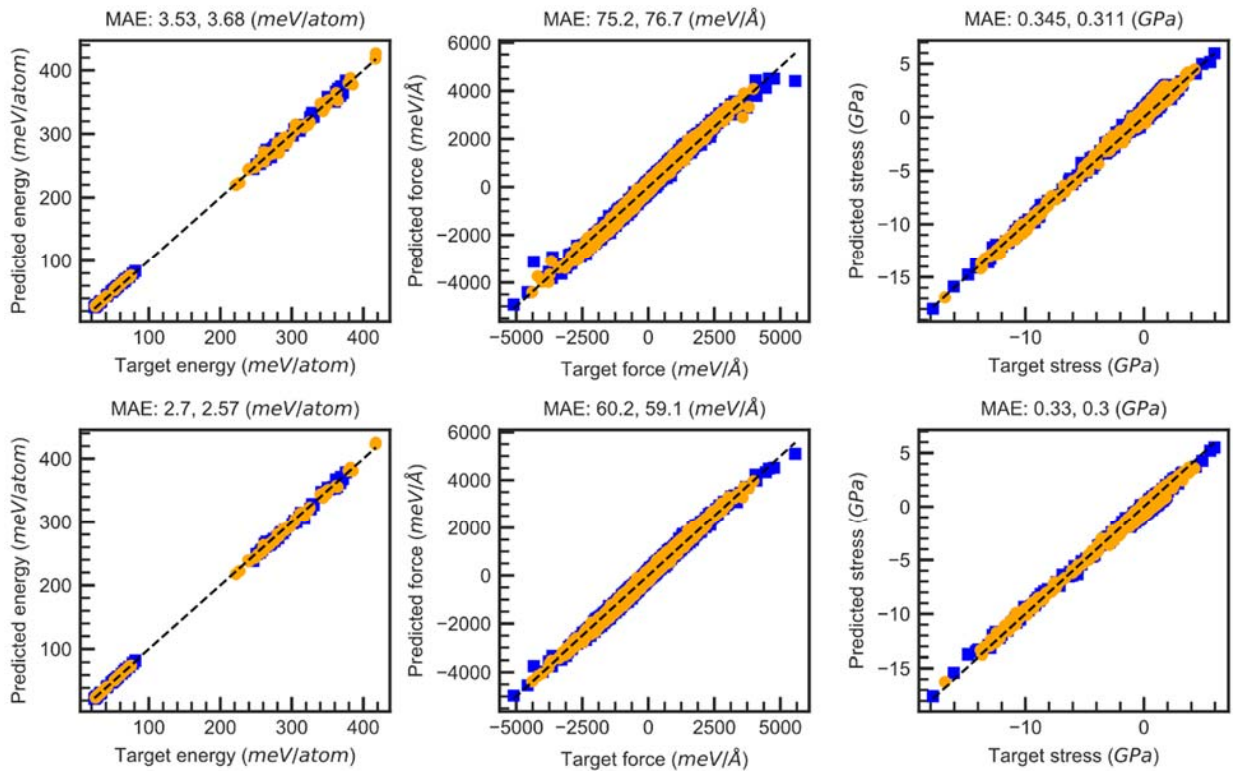

Figure 1. Parity plots of training (orange) and validation (blue) energies, components of force and components of the virial stress tensor for the interatomic potential GP1 (a) and GP2 (b). The black dashed line is the identity. The mean absolute error (MAE) is presented above each sub-figure for validation and training data respectively.

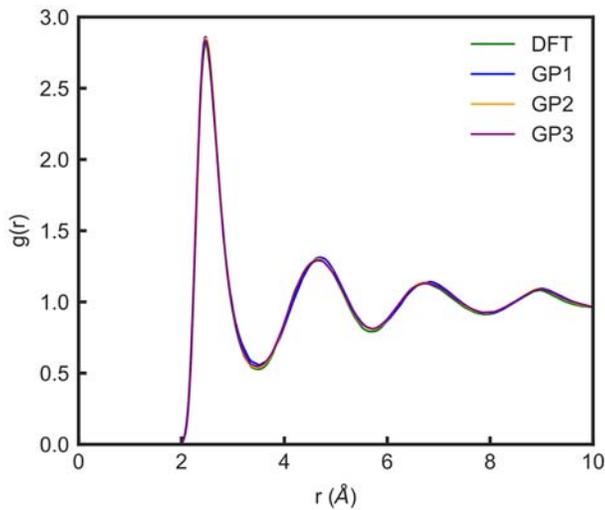

Figure 2. Radial distribution functions of liquid copper at 1400K

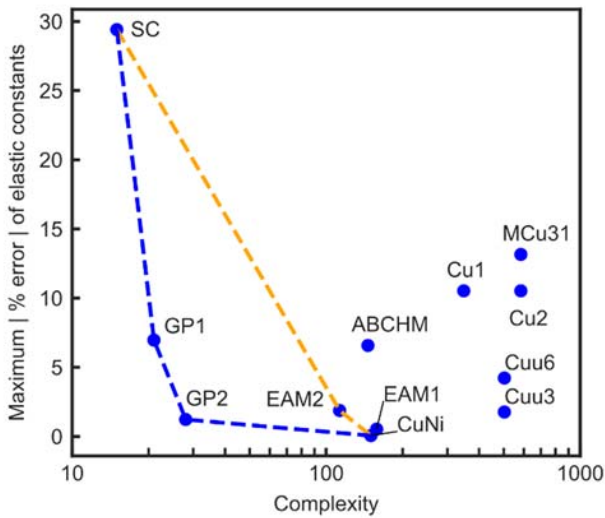

Figure 3. Pareto frontiers of interatomic potentials for copper. No model has less error and is less complex than a model in the Pareto frontier. The orange dashed line was the Pareto frontier before the development of GP1 and GP2, and the blue dashed line is the new Pareto frontier. The percent error for each model was evaluated against the model's own target values, described in the Supplementary Information. Complexity was measured by the number of nodes in the tree representation of the model. Because the smoothing function for some models is unknown, to construct this plot each smoothing function was counted as 2 nodes, representing the smoothing function and a multiplication operation. Sources: SC[54], ABCHM[56], Cu1[56], EAM1[55], EAM2[55], Cu2[59], Cuu6[60], Cuu3[61] and CuNi[57]. The interatomic potentials were found in the Interatomic Potentials Repository.[62] GP1 and GP2 were developed in this work.

Table 3. Error of the values predicted by interatomic potentials for copper relative to the respective reference. The models displayed in this table are near the Pareto frontiers in Figure 3, values of other potentials are in Tables S2 to S7. $C_{ij}$ are elastic constants, $a_0$ is the lattice parameter, $\Delta E$ (bcc-fcc) is the energy difference between bcc and fcc phases, $E_v$ is the fcc bulk vacancy formation energy, $E_{v\ (unrelaxed,\ 2\times2\times2)}$ is the unrelaxed vacancy formation energy computed on a 2×2×2 supercell, $E_m$ is the migration energy for fcc bulk vacancy diffusion, $E_a$ is the activation energy for fcc bulk vacancy diffusion, $E_{dumbbell}$ is the dumbbell <100> formation energy, $\nu$ is the phonon frequency, and $\gamma_{ISF}$ and $\gamma_{USF}$ are the intrinsic and unstable stacking fault energies, respectively.

| Property | Metric | SC | GP1 | GP3 | GP2 | EAM2 | ABCHM | CuNi | EAM1 |
|---|---|---|---|---|---|---|---|---|---|
| Complexity | Number of nodes | 15 | 21 | 26 | 28 | 113 | 146 | 150 | 158 |
| $C_{11}$ | % error | -3.6[a] | 5.8[b] | 2.9[b] | -0.7[b] | 1.9 | -0.6 | 0.1 | -0.1 |
| $C_{12}$ | % error | 3.8[a] | 7.0[b] | 2.5[b] | 0.5[b] | 0.2 | -4.1 | 0.0 | 0.1 |
| $C_{44}$ | % error | -29.4[a] | -2.0[b] | -0.4[b] | -1.2[b] | 0.0 | -6.6 | 0.0 | 0.5 |
| $a_0$ (fcc) | % error | 0.0 | -0.3 | 0.2 | 0.3 | 0.0 | -0.7 | 0.0 | 0.0 |
| $a_0$ (bcc) | % error | - | 0.1 | -0.2 | -0.1 | - | 2.4 | 0.9 | - |
| $\Delta E$ (bcc – fcc) | pred. – ref. (meV/atom) | - | 8 | 12 | 4 | 2[c] | -11 | -13 | 2 |
| $\Delta E$ (hcp – fcc) | pred. – ref. (meV/atom) | - | -3 | -1 | -2 | -6[c] | -2 | -4 | -4 |
| $E_{v\ (unrelaxed,\ 2\times2\times2)}$ | pred. – ref. (meV) | - | 32 | -106 | -123 | - | 80 | - | - |
| $E_v$ | pred. – ref. (meV) | - | 138 | 12 | 2 | -17 | - | 6 | -3 |
| $E_m$ | pred. – ref. (meV) | - | -106 | -49 | -37 | -20 | - | -20 | -21 |
| $E_a = E_v + E_m$ | pred. – ref. (meV) | - | 32 | -36 | -34 | -37 | -24[d] | -14[d] | -24 |
| $E_{dumbbell}$ | pred. – ref. (meV) | - | 49 | -15 | -56 | - | 250 | - | - |
| $\nu_L(X)$ | % error | - | 8.2 | 4.1 | 3.2 | 7.6 | - | 0.0 | 6.0 |
| $\nu_T(X)$ | % error | - | 0.7 | 0.1 | 0.0 | 1.2 | - | -0.2 | 0.8 |
| $\nu_L(L)$ | % error | - | 6.5 | 1.7 | 0.5 | 6.6 | - | -0.8 | 4.6 |
| $\nu_T(L)$ | % error | - | -2.2 | -2.3 | -3.0 | -1.5 | - | -9.4 | -2.6 |
| $\nu_L(K)$ | % error | - | 9.1 | 5.1 | 4.3 | 8.0 | - | -0.8 | 5.4 |
| $\nu_{T1}(K)$ | % error | - | 1.6 | 0.3 | 0.1 | 2.4 | - | 1.1 | 1.1 |
| $\nu_{T2}(K)$ | % error | - | 6.9 | 3.0 | 2.2 | 9.0 | - | 0.9 | 7.0 |
| $\gamma_{ISF}$ | pred. – ref. (mJ/m$^2$) | - | -29 | -6 | -20 | -9 | - | 0 | -1 |
| $\gamma_{USF}$ | pred. – ref. (mJ/m$^2$) | - | -44 | -27 | -31 | - | - | - | - |

Note: properties in orange font were used for training and properties in blue font were not used for training. Properties for which target values are not available are marked with a "-". (a) SC was fit to the bulk modulus. (b) elastic constants were used to select GP1, GP2 and GP3 from the convex hull. (c) fit to ensure that $E_{fcc} < E_{bcc}$ and $E_{fcc} < E_{hcp}$. (d) fit to vacancy formation energy.

There is also good agreement between the newly discovered potential models and other DFT-calculated properties (Table 3). Other models near the Pareto frontiers also show good agreement with their target values, but a notable difference is other than being more complex, these models were also directly trained on many of the properties listed in Table 3 whereas GP1 and GP2 were not. The errors on the elastic constants predicted by GP2 are almost as small as for EAM1, and the simpler model GP1 has errors on elastic constants that are comparable to ABCHM. The GP1 and GP2 models perform well on properties involving hcp and bcc phases, even though no hcp or bcc data were included in the training set. For the bcc lattice constant, the relative energy between the fcc and bcc phases, and the relative energy between fcc and hcp phases, GP1 and GP2 perform comparably to models that were trained on those data points and outperform all models that were not trained on them.

For vacancy formation energies in the dilute limit, GP2 performs very well, with an error of 2 meV relative to the extrapolated DFT energy (see Supplementary Information for details). GP1 performs less well, with an error of 138 meV. Comparisons with other models for vacancy formation energies are difficult, as the models that report their performance on vacancy formation energies were trained with those values, whereas GP1 and GP2 were not. An exception is a neural network potential we discuss later, for which the extrapolated error is 146 meV, comparable to GP1 (Table S6, Supplementary Information). The GP1 error in vacancy formation energy is largely offset by an error in the opposite direction for migration energy, and as a result the errors for both GP1 and GP2 for the activation energy for vacancy-mediated diffusion are comparable to models that were trained on that value.

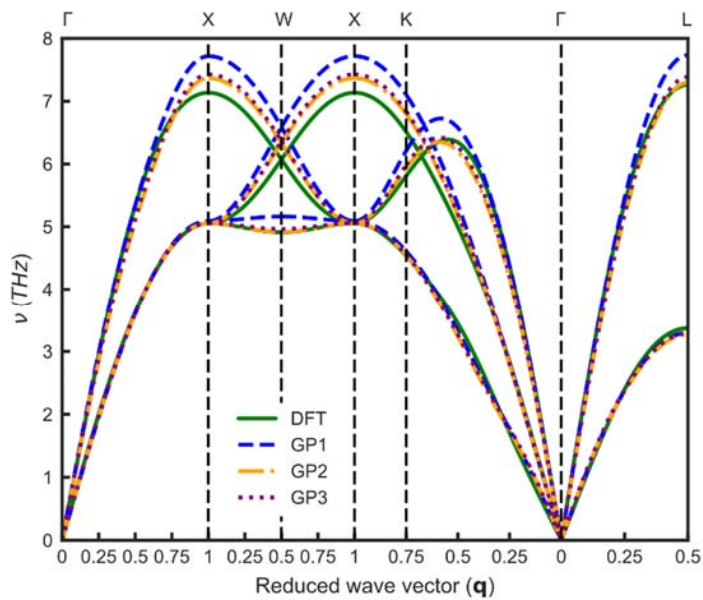

Figure 4. Calculated phonon dispersion curves for DFT, GP1, GP2, and GP3.

GP1 and GP2 also demonstrate good predictive accuracy on phonon frequencies, which were also not included in their training set (Figure 4 and Table 3). On average, GP2 outperforms all other models on phonon frequencies used for validation, with a mean absolute error of 2.0%, and it also outperforms EAM1 on the phonon frequencies on which EAM1 was trained. GP1 does not do as well as GP2 on phonon frequencies, performing on average slightly better than EAM2 but worse than EAM1 and CuNi. The difference in the performance of GP1 and GP2 on phonons is evident in their calculated phonon dispersion curves (Figure 4). The strong performance of GP2 on phonon frequencies and elastic constants suggests that it does well at capturing the curvature of local minima on the potential energy surface, but it may not do as well in states away from the local minima, such as the vacancy formation energy of an unrelaxed 2×2×2 fcc unit cell (Table 3).

Both GP1 and GP2 perform better than the other models for the formation energy of a dumbbell defect. The absolute errors for GP1 and GP2 are only 49 meV and 56 meV respectively, as compared to an absolute prediction error of 250 meV for the ABCHM model (Table 3). Of the three models that included the dumbbell defect formation energy in their training data, the best has an absolute error that is about twice that of the absolute prediction error of GP2 (Table S4). On the other hand, both GP1 and GP2 underestimate the formation energy of a stable intrinsic stacking fault (see Supplementary Information for details) to a greater extent than the other models that report a comparison to this value. The largest absolute error, 29 mJ / m$^2$ for GP1, is 10.2 meV / atom along the (111) plane of the fault. GP1 and GP2 similarly underestimate the formation energy of an unstable stacking fault, but it is hard to assess how this compares to other models as none of the other models reports a benchmark value for the unstable stacking fault energy.

EAM-type models are well known to underpredict surface energies. Surface energies predicted by EAM-type models trained on ab-initio calculations for copper are about 40-50% below their target values for the (100), (110) and (111) surfaces (Table S6). In contrast, GP1 underpredicts these surface energies by only 8%, 1% and 5% respectively, and GP2 underpredicts them by 14%, 10% and 10% respectively (Figure 5). For potentials that use experimental data for their target values, evaluating performance in calculating surface energies is more difficult as only the average value of experimental surface energies is available.[55,57,61] To make this comparison we have calculated weighted average surface energies over 13 different low-index surface facets, where the weights are based on the relative surface areas in Wulff constructions (details are provided in the Supplementary Information)[63]. EAM1 and Cuu3 underpredict the weighted surface energies by about 30%, and CuNi overpredicts the weighted surface energies by about 10% (Table S6).[56] GP1 underpredicts the weighted surface energies by 8% and GP2 by 13%. GP1-predicted surface energies are the most accurate of any of the evaluated EAM-type potential models relative to its target values.

The performance of GP1 and GP2 on surface energies is remarkable because there were no surfaces in the training set; this is a case of machine-learning potential models demonstrating extrapolative predictive ability. Similarly, both GP1 and GP2 demonstrated high predictive accuracy for the dumbbell defect compared to the other models, indicating that they are able to accurately predict energies in both low-coordination and high-coordination environments. There are likely two reasons for the predictive accuracy of these models. The first is that other than SC, GP1 and GP2 are the simplest models considered here, and in general simpler models are less likely to overfit the training data.[49] A similar trend of simpler models demonstrating greater extrapolative ability was observed by Zuo et al. in a recent comparison of different types of machine learned potential models.[19] The second reason is that these models were discovered in a hypothesis space designed to contain models resembling those for which there is fundamental physical justification. In general, the more physics can be included in the machine learning procedure, the more likely it is that a model will have extrapolative predictive power.

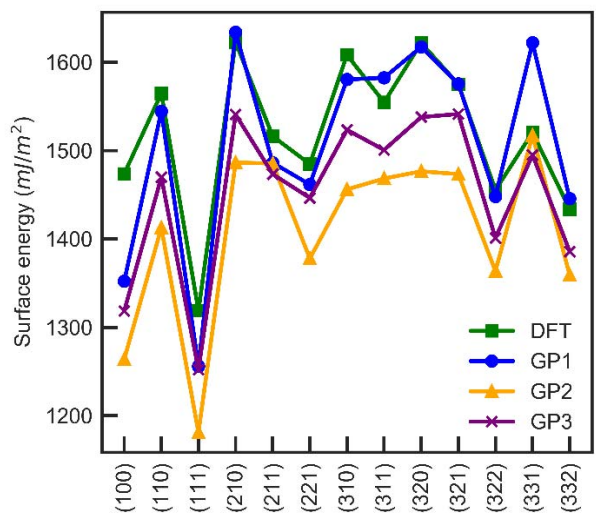

Figure 5. Surface energies of elemental copper as computed using DFT, and the interatomic potentials GP1, GP2, and GP3.

When new data are added to the training set, the genetic programming search for new models can build off of what has been previously learned by using known high-performing models to seed the search. As a demonstration of this approach, we have performed an additional search using an augmented training set in which the 13 low-index surfaces (shown in Figure 5) were added to the training data and always included in the subsets of data used to evaluate candidate models. This search was seeded with GP1 and GP2, and as a result the models it discovered (Table S10) had many features in common with these. One of these models, which we label GP3 (equation (6)) resembles GP2 but, as expected, demonstrates better performance on surface energies (Figure 5). The absolute error for the weighted surface energies is 7% for GP3, as compared to 13% for GP2. The equation for GP3, which is slightly simpler than that of GP2, is provided below.

$$7.51\sum r^{3.98-3.93r} f(r) + \left(28.01 - 0.03\sum r^{11.73-2.93r} f(r)\right)\left(\sum f(r)\right)^{-1} \qquad (6)$$

On average, the improved performance on surface energies for GP3 does not significantly affect its performance on the other properties listed in Table 3 compared to GP2. GP3 performs worse on average on elastic constants and phonon frequencies, but significantly better on the dumbbell formation energy and stacking fault energies. It is difficult to assess the extent to which these changes in performance can be attributed to the addition of surfaces to the training data due to the stochastic nature of the search.

Although GP1, GP2, and GP3 are simpler than many other EAM-type models, they have a similar computational cost when implemented in LAMMPS[29,64] due to the extensive use of tabulated values. Based on our benchmarks (Figure S1) GP1 takes 2.1 μs/step/atom, GP2 3.5 μs/step/atom, and GP3 takes 3.6 μs/step/atom, whereas EAM1 has a cost of 3.0 μs/step/atom. These speeds rank them among the fastest potential models, capable of modeling systems at large time and length scales.[29]

**DISCUSSION**

There are advantages and disadvantages to the different approaches for using machine learning to generate potential models. In many machine learning approaches, including (but not limited to) neural network potentials, Gaussian approximation potentials, moment tensor potentials, SNAP potentials, and AGNI force fields,[1-5] the general idea is to construct a highly flexible hypothesis space that respects local symmetry and, with the help of large amounts of training data, identify the models within that hypothesis space that best reproduce the training data. Such models are capable of achieving very high accuracy for systems in which the local environments of the atoms are similar to those contained in the training set. These machine learning algorithms typically produce potential models that are orders of magnitude faster than DFT but also orders of magnitude slower than EAM-type potentials.[29,53,65-67]

Here we have demonstrated that machine learning can also be used to develop the types of simple, fast potential models that are needed to model systems at extreme time and length scales. The key to our approach is to use genetic programming to search for computationally simple and efficient models in a hypothesis space that is constructed so that it contains simple models that are also physically meaningful. The models are then selected based on a combination of simplicity, speed, and accuracy relative to the training data. The use of simplicity as a selection criterion results in models that are more likely to generalize well, and it also significantly reduces the amount of data required to train the model.[68-70] For example, GP1 and GP2 were trained with 75 32-atom structures, for a total of 2400 atomic environments. For comparison, Artrith and Behler[71] have constructed a neural network potential for copper with a focus on surfaces. The potential was trained using 554,187 atomic environments, including tens of thousands of slabs and cluster structures. It performs comparably to GP1 and GP2 for many bulk properties, and much better for surface energies (Table S6, Supplementary Information). The neural network approach demonstrates very low errors on the types of systems on which it was trained, but as the genetic programming approach requires less training data it is likely that some accuracy can be recovered by using more accurate (and computationally expensive) methods to generate the training data.

The potential models discovered by the genetic programming approach are as fast as EAM-type models and demonstrate good predictive accuracy on properties they were not trained on. In particular, GP1 and GP2 show surprisingly good performance in predicting surface energies (the GP1 mean absolute error for surface energies is only 35mJ/m$^2$) despite the fact that there were no surfaces in their training data. Trained only on DFT data, the genetic programming algorithm found models that resemble widely-used glue potentials with a unique form for the many-body term that depends on the inverse of a sum over pair interactions. One of the advantages of generating potential models using simple analytical expressions is that it may be possible to analyze the expressions to get an insight into the underlying physical interactions that are responsible for the shape of the potential energy surface.

There are some notable limitations and areas for improvement for the approach presented here. For each system studied, it will be necessary to ensure that the hypothesis space contains simple expressions that capture important contributions to the potential energy; for example, for many systems it will likely be necessary to introduce terms that depend on bond angles, which was not done in this work. We used fixed inner and outer cutoff distances in this study, but it would almost certainly be better to let them vary as do other parameters of the potential. There is also the question of how to determine which of the models discovered by the genetic programming algorithm provide the best balance of speed and predictive accuracy. This could be achieved in a number of ways,[40,72,73] including by evaluating performance against validation data, but it is not clear which approach is best. Finally, the genetic programming approach is likely not suitable for on-the-fly learning. Because it is a stochastic method, it can take an indeterminate amount of time to find a set of promising models, and there is no guarantee that an incremental change to the training data will result in an incremental change to the shapes of the potential energy surfaces on the convex hull. Other potential model approaches are probably better-suited for this purpose. Despite these current limitations, our results demonstrate that machine learning holds great promise to improve the accuracy of atomistic calculations at extreme time and length scales.

**METHODS**

**Description of the hypothesis space**

Our machine learning algorithm uses genetic programming to search a hypothesis space of models that can be constructed by combining real numbers, addition, subtraction, multiplication, division, exponentiation, and a sum over neighbors of an atom. As discussed previously in the text, the hypothesis space was based on physical principles. Within this hypothesis space, each function can be represented as a tree graph, as shown in Figure 6. The space was constrained so that the maximum number of summations over neighbors was 6, no nested summations over neighbors were allowed, the maximum allowed depth of a tree was 32 and the maximum allowed number of nodes was 511. To ensure smoothness of the potential, all functions of distances are multiplied by the following smoothing function before the sum over neighbors is taken:[74]

$$f(r) = \left(2r^2 - 3r_{in}^2 + r_{out}^2\right)\left(r_{out}^2 - r^2\right)^2 \left(r_{out}^2 - r_{in}^2\right)^{-3} \qquad (7)$$

where $r_{in}$ and $r_{out}$ are the inner and outer cutoff radii, for GP1 and GP2, $r_{in}$ = 3 Å and $r_{out}$ = 5 Å, including the 3$^{rd}$ nearest neighbors.[55]

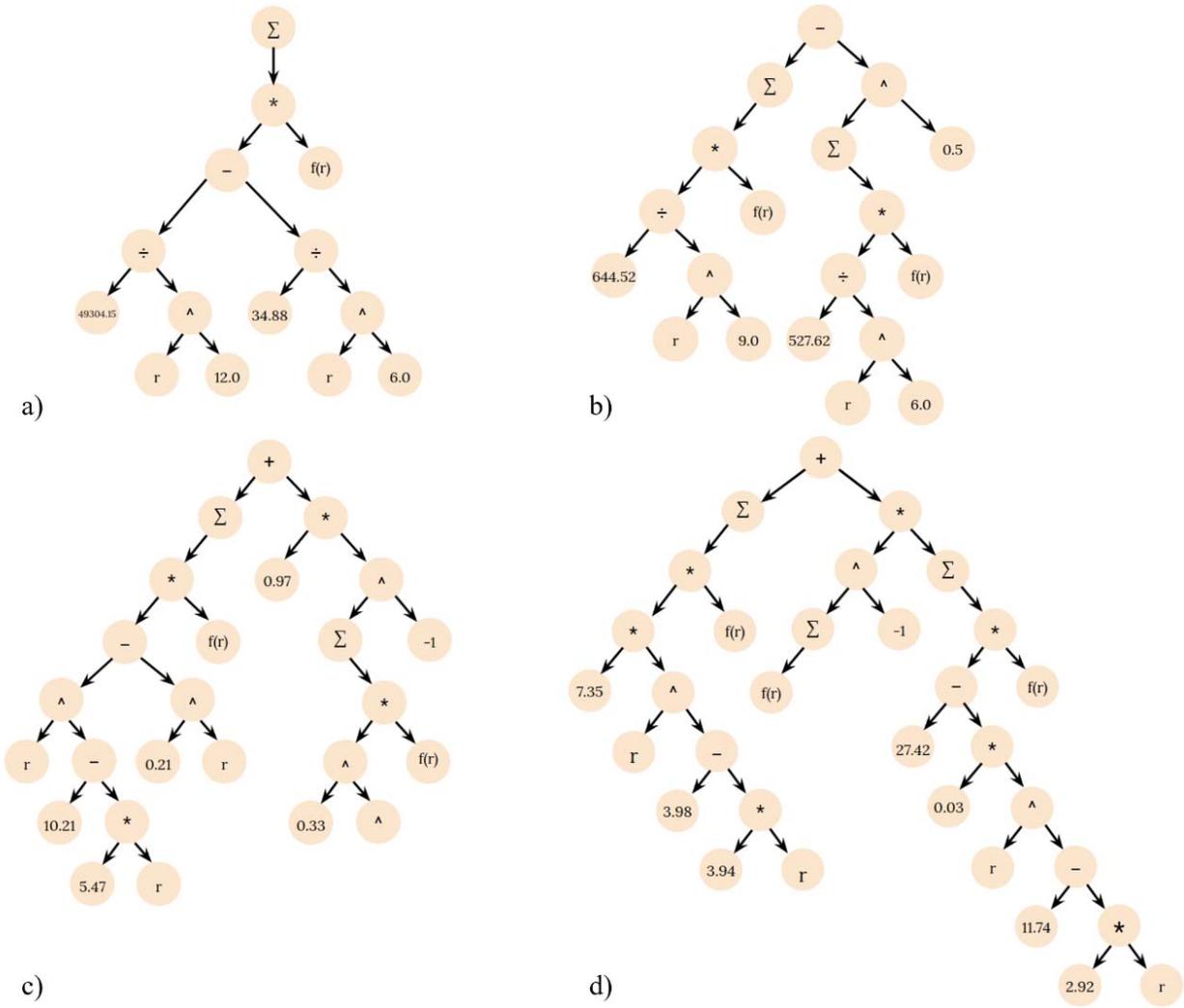

Figure 6. Tree graphs of a) Lennard-Jones potential parametrized for argon, equation (1), b) Sutton-Chen EAM potential parametrized for copper, equation (3), c) GP1 and d) GP2

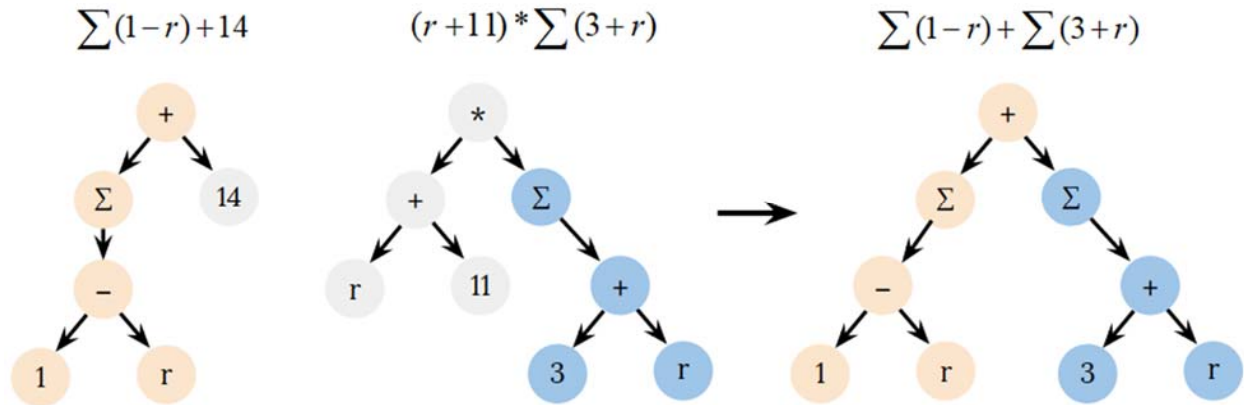

Figure 7. Example of a crossover operation

**Description of the algorithm**

Genetic programming evolves computer programs following Darwin's natural selection by performing crossover and mutation operations on a set of individuals. Crossover was performed by 2 different operations: by randomly selecting a branch from one tree and replacing it with a randomly selected branch of another tree (Figure 7), and by creating a linear combination of 2 randomly selected branches from 2 different tress – the first method was randomly selected 90% of the time and the second one 10% of the time. The mutation operation performed 3 different sub-operations with equal probability: crossover of a tree with a randomly generated tree, swapping the arguments of a binary non-commutative function, and slightly modifying the expression tree by replacing (or inserting) a randomly selected non-terminal node with a randomly selected operator.[75] The randomly generated trees were generated with the grow or full method with equal probability,[36] and the depth was drawn from a Gaussian distribution of mean 5 and standard deviation of 1. The overall algorithm performed crossover with a probability of 0.9, and mutation with a probability of 0.1.

Increasing diversity is known to improve the quality of the optimization.[75] To increase diversity, we implemented a hierarchical way of creating separate environments in which the individuals (i.e., potential models) evolved. We ran the algorithm on 12 processors, and each processor had its own environment, consisting of a population of models and a subset of the training data. Conceptually this allows potentials within a specific environment to develop characteristics that are unique, increasing the diversity. Candidates for crossover and mutation were selected from 3 different sets of models with equal probability:

(1) The population of the current processor. Every 20,000 crossover and mutation operations, 100 individuals were selected based on their fitness (equation (5)) with Pareto tournament selection of size 10 while the rest were discarded.[72,76,77]

(2) A global set of models. Each processor tried to add the 100 individuals selected in part (1) to the global subset every 20,000 crossover and mutation operations. The models on the global set were then evaluated on the basis of speed (to model large time and length scales), fitness (for accurate results), and complexity (for generalizability). The speed of each model was estimated by the number of summations over neighbors. The complexity was evaluated by the number of nodes in the tree graph. To identify the best models, we generated separate convex hulls with respect to fitness and complexity for each number of summations (speed) in a potential. Only the models on these convex hulls were retained in the global set.

(3) Individuals from other processors. Each processor was allowed to communicate with other processors every 5000 crossover and mutation operations, importing the current set of individuals from them.

Selection with equal probability was performed when getting an individual from the global set. Tournament selection of size 10 was used for getting individuals from the population of the current processor and from the populations of other processors.

The training data was also arranged in hierarchical subsets to increase diversity and reduce the speed of evaluating fitness. Globally, a subset of 75 energies, 75 forces, and 75 stresses was randomly sampled from the full set of training data every 20,000 crossover and mutation operations. The fitness of the global set of models was evaluated using this subset of training data. The training data on each processor (15-30 energies, forces and stresses) were randomly selected from the global subset of the training data, and this local subset was used to evaluate fitness locally on each processor. The subset of training data for each processor was selected from the global subset because individuals that migrate from a processor to the global set are more likely to survive if the environment is similar.

Optimization of potential model parameters was performed using the covariance matrix adaptation evolution strategy (CMA-ES) optimizer and a conjugate gradient (CG) optimizer.[78,79] The CMA-ES algorithm was selected because it performs well in nonlinear or non-convex problems. The potential models on the global set of best individuals were optimized with the CMA-ES every 10,000 crossover and mutation operations by one processor. In contrast, the CG algorithm performed one optimization step for every individual generated by crossover or mutation.

The genetic programming algorithm took about 330 CPU-hours to find the exact Lennard-Jones potential, 3600 CPU-hours to find the exact Sutton Chen potential, and 360 CPU-hours to find GP1, GP2 and GP3. We note that it is likely that with additional tuning and performance enhancements the efficiency of the algorithm can be improved. To facilitate this, our code is open source and available at https://gitlab.com/muellergroup/poet.

**Details about the target data**

The DFT data were computed using the Vienna Ab initio Simulation Package[80] (VASP) with the Perdew-Burke-Ernzerhof[81] (PBE) generalized gradient approximation (GGA) exchange correlation functional. The projector augmented wave method[82] (PAW) Cu_pv pseudopotential was used for copper. Efficient $k$-point grids were obtained from the $k$-point grid server with MINDISTANCE = 50Å.[83] A cutoff energy of 750 eV and ADDGRID = TRUE in VASP were required to converge the stress tensor to less than 0.05 GPa. The elastic constants were converged to within 3 |% error| using a MINDISTANCE = 100Å. The DFT point defect energies were computed by linear extrapolation (see Supplementary Information for more details). The phonon dispersion curves were computed on a 3×3×3 supercell. The DFT calculation used a 5×5×5 $k$-point grid and electronic self-consistency convergence of $10^{-8}$ eV. The radial distribution function molecular dynamics simulations were performed in the NVT ensemble at the experimental 1400K liquid density on a 3×3×3 supercell. The temperature was increased from 300K to 2500K during 1 ps. Then the temperature was maintained at 2500K during 10 ps. Then, the temperature was decreased from 2500K to 1400K over 1ps. Then the temperature was maintained at 1400K during 1ps. Finally, the radial distribution function data was collected at 1400K over 40 ps. The DFT molecular dynamics for the radial distribution function was performed with a cutoff energy of 400 eV for the equilibration steps and 750 eV for the final 40 ps during which data was collected. Electronic self-consistency convergence was $10^{-5}$ eV and only the $k$-point at Γ was used. For the computation of the fitness of the models, the energies were transformed by subtracting the minimum and dividing by the standard deviation, and the forces and stresses were standardized by subtracting the mean and dividing by the standard deviation.

The data used to rediscover the Lennard-Jones potential and the SC potential, and the data used to validate GP1 and GP2 were computed on LAMMPS. Instructions and files required to use GP1 and GP2 on LAMMPS are provided on the Supplementary Information. Lennard-Jones calculations used a cutoff distance of 7.5 Å, and SC, GP1, GP2 and GP3 calculations used a cutoff distance of 5 Å.

**Benchmarking model speed**

Benchmarking of model speed was done on a single core of a Haswell node with a clock speed of 2.5 GHz. The benchmarking simulation consisted of 10,000,000 molecular dynamics steps for a 32-atom unit cell.

## DATA AVAILABILITY

Our code is open source and available at https://gitlab.com/muellergroup/poet. The instructions and files required to use GP1, GP2 and GP3 on LAMMPS are provided in the Supplementary Information. The data used to train the models is provided in the Supplementary Information.

## ACKNOWLEDGEMENTS

This work was done using high-performance computing resources from the Maryland Advanced Research Computing Cluster (MARCC) and from the Homewood High-Performance Cluster (HHPC). The authors thank Qing-Jie Li, Zhao Fan, Dihui Ruan and the members of the Mueller Research Group for useful discussions.

## COMPETING INTERESTS

The authors declare no competing financial interests.

## CONTRIBUTIONS

T.M. conceived of and managed the project. T.M. and A.H. developed the software and wrote the manuscript. A.H., A.B. and F.Y. computed the data. A.H. and A.B. worked on open sourcing the software. A.H. and S.M. ran experiments. A.B. implemented the models in LAMMPS. A.H. developed data analysis scripts. All the authors proposed, discussed or developed ideas that improved the performance of the machine learning algorithm or the quality of the data.

## FUNDING

The authors acknowledge financial support from the Office of Naval Research, grant number N000141512665.

**FIGURE LEGENDS**

Figure 1. Parity plots of training (orange) and validation (blue) energies, components of force and components of the virial stress tensor for the interatomic potential GP1 (a) and GP2 (b). The black dashed line is the identity. The mean absolute error (MAE) is presented above each sub-figure for validation and training data respectively.

Figure 2. Radial distribution functions of liquid copper at 1400K

Figure 3. Pareto frontiers of interatomic potentials for copper. No model has less error and is less complex than a model in the Pareto frontier. The orange dashed line was the Pareto frontier before the development of GP1 and GP2, and the blue dashed line is the new Pareto frontier. The percent error for each model was evaluated against the model's own target values, described in the Supplementary Information. Complexity was measured by the number of nodes in the tree representation of the model. Because the smoothing function for some models is unknown, to construct this plot each smoothing function was counted as 2 nodes, representing the smoothing function and a multiplication operation. Sources: SC54, ABCHM56, Cu156, EAM155, EAM255, Cu259, Cuu660, Cuu361 and CuNi57. The interatomic potentials were found in the Interatomic Potentials Repository.[62] GP1 and GP2 were developed in this work.

Figure 4. Calculated phonon dispersion curves for DFT, GP1, GP2, and GP3.

Figure 5. Surface energies of elemental copper as computed using DFT, and the interatomic potentials GP1, GP2, and GP3.

Figure 6. Tree graphs of a) Lennard-Jones potential parametrized for argon, equation (1), b) Sutton-Chen EAM potential parametrized for copper, equation (3), c) GP1 and d) GP2

Figure 7. Example of a crossover operation

**SUPPLEMENTARY INFORMATION**

The Supplementary Information shows tables of the errors of the predictions of the interatomic potentials on: elastic constants, lattice parameters, energy difference between phases, vacancy formation energies, migration and activation energies, and surface energies. Figure S1 shows the computational cost of LJ, SC, GP1, GP2, GP3 and EAM1 as implemented in LAMMPS. Figure S2 shows that the Pareto frontier of absolute percent error on elastic constants against complexity does not change when using the average instead of the maximum error. The Supplementary Information includes the instructions and the files required for enabling GP1, GP2 and GP3 in LAMMPS.

# Supplementary information

# Fast, accurate, and transferable many-body interatomic potentials by symbolic regression

Alberto Hernandez, Adarsh Balasubramanian, Fenglin Yuan, Simon Mason, and Tim Mueller

**Description of the potential models**

*Table S1. Acronyms used for the interatomic potential models*

| Acronym | Description | Fitting procedure |
|---|---|---|
| SC[1] | Sutton-Chen (SC) EAM interatomic potential. Defined by a Finnis-Sinclair[2] potential with a van der Waals term for long-range interactions. Developed for metallic bonding and mechanical interactions between clusters | Experimental data |
| GP1, GP2 and GP3 | Interatomic potential developed in this work | *Ab initio* data |
| EAM2[3] | Compared against EAM1 in the article.[3] Defined by a Morse[4] function, the universal equation of state[5] and the density related to a 4s orbital. Developed for energies and mechanical stability of non-equilibrium structures | Experimental data |
| ABCHM[6] | Extended the potential by Ackland et al.[7] (a Finnis-Sinclair potential[2]) by adding a quadratic term in the embedding function to improve the melting point. Developed for crystallization kinetics from deeply undercooled melts | *Ab initio* and experimental data. Melting properties and crystalline properties |
| CuNi[8] | Defined by a Morse function[4], the universal equation of state[5] and a hyperbolic tangent which considers the shape of the 3d orbital. Developed for Cu-Ni alloys | Experimental data to train the Cu potential |
| EAM1[3] | Based on Morse potentials, unit step functions, and other terms. Widely used, especially for defect properties in copper.[9] Developed for energies and mechanical stability of non-equilibrium structures | Experimental data with *ab-initio* data for relative energies of hcp, bcc, and fcc phases, and energies of the fcc phase and a diatomic molecule under strong compression |
| Cu1[6] | Developed for crystallization kinetics from deeply undercooled melts.[9] | *Ab initio* and experimental data. Crystalline, liquid and melting point data |
| Cuu6[10] | Developed and defined in the same way as Cuu3 but used more accurate vacancy formation energies | Experimental data |
| Cuu3[11] | Defined by the universal equation of state[5] and the spherically averaged free-atom densities calculated from Hartree-Fock theory.[12,13] Developed for several fcc metals and alloys | Experimental data |
| Cu2[9] | Developed in the same way as Cu1, but added twin boundary energy to training data | *Ab initio* and experimental data. Melting properties, crystalline properties, twin boundary energy |
| MCu31[14] | The paper developed several EAM potentials to study the effects of stacking fault energy on dislocation nucleation[(a)] | *Ab initio* and experimental data. Melting properties, crystalline properties, twin boundary energy and stacking fault energy |

Notes: (a) MCu31 was chosen for comparison here because it calculates stacking fault energies more accurately than the others.[15] The equation for MCu31 was not provided, but it was described as being based on Cu2 and fit to additional training data, so we have assigned it the same complexity as Cu2. The acronyms of the models are taken from the original paper or adapted from the Interatomic Potentials Repository.[15]

**Models that trained with ab-initio and experimental data**

ABCHM, Cu1, Cu2, MCu31 and EAM1 used experimental and *ab initio* data for training. About 40 % of the training data of ABCHM, Cu1, Cu2 and MCu31 was *ab initio* data; it included relative energies between phases, lattice parameters, vacancy formation energy and interstitial formation energy. EAM1 used *initio* data for relative energies of hcp, bcc, and fcc phases, and energies of the fcc phase and a diatomic molecule under strong compressions.

**Comparison of GP1, GP2 and GP3 against EAM-type potentials**

The errors of each of the interatomic potentials are obtained directly from the original paper in which the potential was first published, with one exception: the weighted surface energies of CuNi, EAM1 and Cuu3 were computed using the interatomic potentials from the Interatomic Potentials Repository using LAMMPS.[15,17]

*Table S2. Errors on elastic constants and lattice parameters*

| Model | Complexity | C11 | C12 | C44 | a0 (FCC) | a0 (BCC) |
|---|---|---|---|---|---|---|
| Description | Number of nodes | Error | Error | Error | Error | Error |
| Units | Count | % | % | % | % | % |
| SC | 15 | -3.6[a, b] | 3.8[a, b] | -29.4[a, b] | 0.0[a] | - |
| GP1 | 21 | 5.8[c] | 7.0[c] | -2.0[c] | -0.3[c] | 0.1[c] |
| GP3 | 26 | 2.9[c] | 2.5[c] | -0.4[c] | 0.2[c] | -0.2[c] |
| GP2 | 28 | -0.7[c] | 0.5[c] | -1.2[c] | 0.3[c] | -0.1[c] |
| EAM2 | 113 | 1.9[a] | 0.2[a] | 0.0[a] | 0.0[a] | - |
| ABCHM | 146 | -0.6[a] | -4.1[a] | -6.6[a] | -0.7[c] | 2.4[c] |
| CuNi | 150 | 0.1[a] | 0.0[a] | 0.0[a] | 0.0[a] | 0.9[c] |
| EAM1 | 158 | -0.1[a] | 0.1[a] | 0.5[a] | 0.0[a] | - |
| Cu1 | 348 | 2.9[a] | 4.1[a] | 10.5[a] | 0.0[c] | 0.0[c] |
| Cuu6 | 503 | -1.2[a] | 1.2[a] | 4.2[a] | 0.0[a] | - |
| Cuu3 | 503 | -1.8[a] | 1.2[a] | 0.3[a] | 0.0[a] | - |
| Cu2 | 584 | 2.4[a] | 3.3[a] | 10.5[a] | 0.0[c] | 0.0[c] |
| MCu31 | 584 | -1.2[a] | 6.5[a] | 13.2[a] | 0.0[c] | - |

Notes: properties in orange were used for training and properties in blue were used for validation. (a) experiment target data. (b) fit to bulk modulus. (c) *ab initio* calculation target data.

*Table S3. Errors on difference between energies of FCC, BCC and HCP phases*

| Model | Complexity | ΔE(BCC-FCC) | ΔE(HCP-FCC) |
|---|---|---|---|
| Description | Number of nodes | Pred.-Targ. | Pred.-Targ. |
| Units | Count | meV/atom | meV/atom |
| GP1 | 21 | 8 | -3 |
| GP3 | 26 | 12 | -1 |
| GP2 | 28 | 4 | -2 |
| EAM2 | 113 | 2[a] | -6[a] |
| ABCHM | 146 | -11 | -2 |
| CuNi | 150 | -13 | -4 |
| EAM1 | 158 | 2 | -4 |
| Cu1 | 348 | 5 | 0 |
| Cu2 | 584 | 6 | 1 |
| MCu31 | 584 | 0 | -6 |

Notes: properties in orange were used for training and properties in blue were used for validation. All values are from *ab initio* calculations. (a) EAM2 was fit subject to the requirement that the fcc structure be more stable than bcc or hcp.

*Table S4. Errors on bulk vacancy formation energy, migration energy, activation energy and dumbbell <100> formation energy*

| Model | Complexity | $E_{v\ (unrelaxed, 2\times2\times2)}$ | $E_v$ | $E_m$ | $E_a$ | $E_{dumbbell}$ |
|---|---|---|---|---|---|---|
| Description | Num. of nodes | Pred.-Targ. | Pred.-Targ. | Pred.-Targ. | Pred.-Targ. | Pred.-Targ. |
| Units | Count | meV | meV | meV | meV | meV |
| GP1 | 21 | 32[a] | 138[a] | -106[a] | 32[a] | 49[a] |
| GP3 | 26 | -106[a] | 12[a] | -49[a] | -36[a] | -15[a] |
| GP2 | 28 | -123[a] | 2[a] | -37[a] | -34[a] | -56[a] |
| EAM2 | 113 | - | -17[b] | -20[b] | -37[b] | - |
| ABCHM | 146 | 80[a] | - | - | -24[b, c] | 250[a] |
| CuNi | 150 | - | 6[b] | -20[b] | -14[b, c] | - |
| EAM1 | 158 | - | -3[b] | -21[b] | -24[b] | - |
| Cu1 | 348 | -9[a] | - | - | -2[b, c] | -120[a] |
| Cuu6 | 503 | - | 30[b] | -20[b] | -50[b, c] | - |
| Cuu3 | 503 | - | -20[b] | -40[b] | -120[b, c] | - |
| Cu2 | 584 | 53[a] | - | - | -26[b, c] | -120[a] |
| MCu31 | 584 | - | - | - | - | -110[a] |

Notes: properties in orange were used for training and properties in blue were used for validation. (a) *ab initio* target data. (b) experimental target data. (c) fitted to the vacancy formation energy

The dumbbell was formed by displacing an atom along a <100> direction. The relaxed vacancy formation energy, the migration energy, the activation energy, and the dumbbell formation energy of GP1, GP2, GP3, EAM2, and EAM1 were computed with 6x6x6 supercells, CuNi used a supercell with 1200 atoms, Cuu3 500 atoms, and Cuu6 250 atoms. The DFT target values were obtained by linear extrapolation of the values at 2×2×2 and 3×3×3 supercells with respect to the inverse of the supercell size.[18] The dumbbell formation energies of ABCHM, Cu1, Cu2, and MCu31 were computed with a 3×3×3 supercell.

*Table S5 Errors on phonon frequencies*

| Model | Complexity | $v_L(X)$ | $v_T(X)$ | $v_L(L)$ | $v_T(L)$ | $v_L(K)$ | $v_{T1}(K)$ | $v_{T2}(K)$ |
|---|---|---|---|---|---|---|---|---|
| Description | Number of nodes | Error | Error | Error | Error | Error | Error | Error |
| Units | Count | % | % | % | % | % | % | % |
| GP1 | 21 | 8.2[a] | 0.7[a] | 6.5[a] | -2.2[a] | 9.1[a] | 1.6[a] | 6.9[a] |
| GP3 | 26 | 4.1[a] | 0.1[a] | 1.7[a] | -2.3[a] | 5.1[a] | 0.3[a] | 3.0[a] |
| GP2 | 28 | 3.2[a] | 0.0[a] | 0.5[a] | -3.0[a] | 4.3[a] | 0.1[a] | 2.2[a] |
| EAM2 | 113 | 7.6[b] | 1.2[b] | 6.6[b] | -1.5[b] | 8.0[b] | 2.4[b] | 9.0[b] |
| CuNi | 150 | 0.0[b] | -0.2[b] | -0.8[b] | -9.4[b] | -0.8[b] | 1.1[b] | 0.9[b] |
| EAM1 | 158 | 6.0[b] | 0.8[b] | 4.6[b] | -2.6[b] | 5.4[b] | 1.1[b] | 7.0[b] |

Notes: properties in orange were used for training and properties in blue were used for validation. (a) *ab initio* target data. (b) experimental target data.

*Table S6. Prediction errors for surface energies*

| Model | Complexity | %error of weighted surface energy | % error | % error | % error | Mean absolute % error. |
|---|---|---|---|---|---|---|
| Description | Number of nodes | 13 surfaces | (100) | (110) | (111) | 13 surfaces |
| Units | Count | % | % | % | % | % |
| GP1 | 21 | -7.6[a] | -8.2[a] | -1.3[a] | -4.8[a] | 2.3[a] |
| GP3 | 26 | -7.4[a] | -10.5[a] | -6.1[a] | -5.1[a] | 4.4[a] |
| GP2 | 28 | -12.6[a] | -14.2[a] | -9.7[a] | -10.5[a] | 7.2[a] |
| ABCHM | 146 | - | -49.7[a] | -47.4[a] | -53.8[a] | - |
| CuNi | 150 | 9.8[b] | - | - | - | - |
| EAM1 | 158 | -28.4[b] | - | - | - | - |
| Cu1 | 348 | - | -50.0[a] | -48.4[a] | -53.8[a] | - |
| Cuu3 | 503 | -31.8[b] | - | - | - | - |
| MCu31 | 584 | - | -39.3[a] | -37.7[a] | -40.1[a] | - |

Notes: properties in orange were used for training and properties in blue were used for validation. (a) *ab initio* target data. (b) experimental target data.

The weighted surface energy is:[19]

$$\bar{\gamma} = \frac{\sum_{\{hkl\}} \gamma_{hkl} A_{hkl}}{\sum A_{hkl}}$$

Here, γ is the surface energy, $A_{hkl}$ is the total area of all the planes in the {*hkl*} family in the Wulff construction.[20] The % error of the weighted surface energy was computed by comparing the experimental (except for GP1 and GP2) target value reported in the paper against the weighted surface energy predicted by the potential.[19,21] The target value of GP1 and GP2 was the weighted surface energy computed by DFT.

*Table S7 Prediction errors for the intrinsic stacking fault ($\gamma_{ISF}$) energy and the unstable stacking fault ($\gamma_{USF}$) energy*

| Model | Complexity | pred. – ref. (mJ/m$^2$) | pred. – ref. (mJ/m$^2$) |
|---|---|---|---|
| Description | Number of nodes | $\gamma_{ISF}$ | $\gamma_{USF}$ |
| Units | Count | % | % |
| GP1 | 21 | -29[a] | -44[a] |
| GP3 | 26 | -6[a] | -27[a] |
| GP2 | 28 | -20[a] | -31[a] |
| EAM2 | 113 | -9[b] | - |
| CuNi | 150 | 0[b] | - |
| EAM1 | 158 | -1[b] | - |
| MCu31 | 584 | 6[b] | 1[a] |

Notes: properties in orange were used for training and properties in blue were used for validation. (a) *ab initio* target data. (b) experimental target data.

The DFT, GP1, GP2 and GP3 intrinsic stacking fault energy and unstable stacking fault energy were computed with a (111) slab, with a gap between periodic slabs of 20 Å and a thickness of 22 (111) atomic layers. The atoms were only allowed to relax in the direction normal to the slab interface for the USF computation. The intrinsic stacking fault and the unstable stacking fault were formed by displacing atoms above a $\{111\}$ plane along a $\langle 211 \rangle$ direction by $a_0/6$, and $a_0/12$, respectively.

## Comparison of GP1, GP2, and GP3 against neural network potential

*Table S8. Comparison between genetic programming potentials and a neural network potential.[22]*

| | Details | GP1 | GP3 | GP2 | Neural network |
|---|---|---|---|---|---|
| Parameters | Count | 5 | 7 | 8 | 2521 |
| Atomic environments in training set | Count | 2400 | 2651 | 2400 | 554,187 |
| Energy MAE | meV/atom | 3.5 | 2.5 | 2.7 | 2.2 |
| Force MAE | meV/Å | 75 | 62 | 60 | 56 |
| $C_{11}$ | % error | 5.8 | 2.9 | -0.7 | 2.3 |
| $C_{12}$ | % error | 7.0 | 2.5 | 0.5 | -3.3 |
| $C_{44}$ | % error | -2.0 | -0.4 | -1.2 | 3.8 |
| $a_0$ (fcc) | % error | -0.3 | 0.2 | 0.3 | 0.0 |
| $a_0$ (bcc) | % error | 0.1 | -0.2 | -0.1 | 0.1 |
| $\Delta E$ (bcc-fcc) | pred. - ref. (meV/atom) | 8 | 12 | 4 | -4 |
| $\Delta E$ (hcp-fcc) | pred. - ref. (meV/atom) | -3 | -1 | -2 | -7 |
| $E_v$ | pred. - ref. (meV) | 136 | 10 | 0 | 146[a] |
| $E_{v \text{ (relaxed, 3×3×3)}}$ | pred. - ref. (meV) | 111 | -15 | -26 | 106 |
| $E_{v \text{ (relaxed, 2×2×2)}}$ | pred. - ref. (meV) | 53 | -75 | -88 | 10 |
| $E_{v \text{ (unrelaxed)}}$ | pred. - ref. (meV) | 142 | 2 | -15 | 147[a] |
| $E_{v \text{ (unrelaxed, 3×3×3)}}$ | pred. - ref. (meV) | 109 | -31 | -47 | 103 |
| $E_{v \text{ (unrelaxed, 2×2×2)}}$ | pred. - ref. (meV) | 32 | -106 | -123 | -1 |
| (100) surface energy | % error | -8.2 | -10.5 | -14.2 | 0.5 |
| (110) surface energy | % error | -1.3 | -6.1 | -9.7 | 1.5 |
| (111) surface energy | % error | -4.8 | -5.1 | -10.5 | -0.4 |

Notes: properties in orange were used for training and properties in blue were used for validation. All the target properties were computed by DFT. The DFT dilute vacancy formation energies were obtained by linearly extrapolating the values at 2×2×2 and 3×3×3 with respect to the inverse of the supercell size.[18] The dilute values for GP1, GP2, and GP3 were computed with a 6x6x6 supercell. (a) The dilute values for the neural network were computed by extrapolating the values at 2×2×2 and 3×3×3.

# Convex hull of models found by the machine learning algorithm

*Table S9. Errors on different properties for models on the 3-dimensional convex hull in Table 1 of the main text, listed in the order they appear in the table. $C_{ij}$ are elastic constants, $a_0$ is the lattice parameter, $\Delta E$ (bcc-fcc) is the energy difference between bcc and fcc phases, $E_v$ is the fcc bulk vacancy formation energy, $E_{v\ (unrelaxed,\ 2\times2\times2)}$ is the unrelaxed vacancy formation energy computed on a 2×2×2 supercell, $E_m$ is the migration energy for fcc bulk vacancy diffusion, $E_a$ is the activation energy for fcc bulk vacancy diffusion, $E_{dumbbell}$ is the dumbbell <100> formation energy, v is the phonon frequency, $\gamma_{ISF}$ and $\gamma_{USF}$ are the intrinsic and unstable stacking fault energies, respectively, $\bar{\gamma}$ is the average surface energy weighted according to the Wulff construction, and $\gamma_{abs}$ is the mean absolute surface energy over 13 surfaces.*

| Property | Metric | M1 | M2 | M3 | M4 | M5 | M6 | GP1 | M8 | M9 | GP2 | M11 |
|---|---|---|---|---|---|---|---|---|---|---|---|---|
| Complexity[a] | Number of nodes | 2 | 4 | 8 | 10 | 13 | 15 | 21 | 25 | 19 | 28 | 29 |
| $C_{11}$[a] | % error | 4081.6 | -85.7 | 5.1 | 24.6 | 40.2 | 9.1 | 5.8 | 5.8 | 28.2 | -0.7 | 2.3 |
| $C_{12}$[a] | % error | 3827.2 | -84.5 | -35.7 | -17.4 | 21.1 | 12.9 | 7.0 | 7.0 | 24.8 | 0.5 | 1.6 |
| $C_{44}$[a] | % error | 4872.1 | -91.7 | 3.3 | 32.5 | 10.3 | -5.5 | -2.0 | -1.9 | 7.1 | -1.2 | -3.6 |
| $a_0$ (fcc) | % error | - | 95.5 | -0.7 | -1.4 | -0.5 | -0.2 | -0.3 | -0.3 | -0.2 | 0.3 | 0.2 |
| $a_0$ (bcc) | % error | 23.0 | 100.8 | 7.5 | -2.4 | -1.2 | 0.2 | 0.1 | 0.0 | -0.6 | -0.1 | -0.2 |
| $\Delta E$ (bcc – fcc) | pred. – ref. (meV/atom) | 43026 | -36 | 74 | 97 | 89 | 8 | 8 | 8 | 42 | 4 | 12 |
| $\Delta E$ (hcp – fcc) | pred. – ref. (meV/atom) | -5 | -5 | 18 | 13 | 12 | -3 | -3 | -3 | 3 | -2 | -2 |
| $E_{v\ (unrelaxed,\ 2\times2\times2)}$ | pred. – ref. (meV) | -1117 | -1117 | 167 | 224 | 66 | 41 | 32 | 33 | -71 | -123 | -207 |
| $E_v$ | pred. – ref. (meV) | -970 | -970 | 104 | 249 | 76 | 138 | 138 | 33 | 21 | 2 | -89 |
| $E_m$ | pred. – ref. (meV) | 10054 | -709 | -286 | -79 | -157 | -133 | -106 | 139 | -72 | -37 | 42 |
| $E_a$ | pred. – ref. (meV) | 9084 | -1679 | -182 | 170 | -81 | 6 | 32 | -106 | -51 | -34 | -47 |
| $E_{dumbbell}$ | pred. – ref. (meV) | $-6\times10^{24}$ | -2925 | 99 | 737 | 372 | -28 | 49 | 33 | 220 | -56 | -41 |
| $v_L(X)$ | % error | 206.3 | -89.5 | 10.5 | 21.1 | 12.8 | 7.0 | 8.2 | 8.3 | 7.4 | 3.2 | 1.9 |
| $v_T(X)$ | % error | 205.9 | -89.5 | -5.4 | 9.8 | 1.5 | -0.9 | 0.7 | 0.9 | 2.0 | 0.0 | -0.8 |
| $v_L(L)$ | % error | 200.9 | -89.7 | 13.2 | 22.1 | 13.6 | 5.4 | 6.5 | 6.5 | 6.2 | 0.5 | -0.7 |
| $v_T(L)$ | % error | 223.6 | -88.9 | 0.8 | 15.3 | 6.5 | -3.9 | -2.2 | -2.0 | 2.6 | -3.0 | -2.0 |
| $v_L(K)$ | % error | 209.9 | -89.4 | 9.4 | 20.7 | 12.5 | 7.8 | 9.1 | 9.1 | 8.0 | 4.3 | 2.9 |
| $v_{T1}(K)$ | % error | 214.5 | -89.2 | -4.1 | 11.3 | 3.2 | 0.0 | 1.6 | 1.8 | 3.0 | 0.1 | 0.1 |
| $v_{T2}(K)$ | % error | 211.0 | -89.4 | 8.7 | 20.0 | 11.7 | 6.2 | 6.9 | 7.6 | 6.8 | 2.2 | 1.4 |
| $\gamma_{ISF}$ | pred. – ref. (mJ/m$^2$) | -90 | -46 | 188 | 158 | 138 | -27 | -29 | -29 | 44 | -20 | -19 |
| $\gamma_{USF}$ | pred. – ref. (mJ/m$^2$) | -216 | -172 | 42 | 88 | 30 | -49 | -44 | -44 | -2 | -31 | -29 |
| $\bar{\gamma}$ | % error | -100.0 | -100.0 | -1.4 | 11.3 | 41.6 | -3.0 | -7.6 | -7.6 | 2.1 | -12.6 | -19.1 |
| $\gamma_{abs}$ | \| % error \| | 100.0 | 100.0 | 7.3 | 7.9 | 37.3 | 3.8 | 2.3 | 2.3 | 3.9 | 7.2 | 16.2 |
| $\gamma_{(100)}$ | % error | -100.0 | -100.0 | -2.1 | 0.9 | 26.8 | -3.7 | -8.2 | -8.2 | -1.7 | -14.2 | -21.7 |
| $\gamma_{(110)}$ | % error | -100.0 | -100.0 | -7.3 | 6.2 | 29.6 | 3.9 | -1.3 | -1.2 | 2.1 | -9.7 | -17.4 |
| $\gamma_{(111)}$ | % error | -100.0 | -100.0 | 20.2 | 37.7 | 50.2 | 0.0 | -4.8 | -4.7 | 4.4 | -10.5 | -17.0 |

Notes: (a) This property was used for model selection from the convex hull.

*Table S10. The 3-dimensional convex hull of models found by seeding with GP1 and GP2 and including the 13 low-index surfaces in the training data*

| Fitness | Cost* | Complexity | Expression |
|---|---|---|---|
| 5404837 | 1 | 2 | $\sum rf(r)$ |
| 1785.2 | 1 | 4 | $-55.16\left(\sum rf(r)\right)^{-1}$ |
| 98.18 | 1 | 8 | $\sum (669.45r^{-9.83} - 0.10)f(r)$ |
| 58.56 | 1 | 10 | $\sum (r^{10.21-5.48r} - 0.07)f(r)$ |
| 55.35 | 4 | 13 | $\left(1431.13\left(\sum f(r)\right)\left(\sum r^{-12.95}f(r)\right)\right)^{0.02\sum f(r)} - 0.12\sum f(r)$ |
| 10.54 | 2 | 15 | $\sum r^{10.21-5.47r}f(r) + 0.98\left(\sum 0.31^r f(r)\right)^{-1}$ |
| 9.08 | 2 | 21 | $2.74\sum r^{7.36-4.86r}f(r) + 29.59\left(\sum r^{3.61-1.19r}f(r)\right)^{-1}$ |
| 9.03 | 2 | 25 | $\sum \left(2.75r^{7.36-4.86r} - 0.0009\right)f(r) + 29.73\left(\sum r^{3.61-1.19r}f(r)\right)^{-1} - 15.98$ |
| 5.98 | 3 | 26 | $7.51\sum r^{3.98-3.93r}f(r) + \left(28.01 - 0.03\sum r^{11.73-2.93r}f(r)\right)\left(\sum f(r)\right)^{-1}$ |
| 5.97 | 4 | 31 | $\sum \left(7.50r^{3.98-3.93r} + 0.001\right)f(r) + \left(28.5 - 0.03\sum r^{11.73-2.94r}f(r)\right)\left(\sum f(r)\right)^{-1}$ |
| 5.54 | 4 | 39 | $7.03\sum r^{4.00-3.88r}f(r) + \left(26.18 - 0.03\sum r^{11.73-2.94r}f(r)\right)\left(\sum f(r)\right)^{-1} + 0.92\left(\sum r^{3.50-1.52r}f(r)\right)^{-1} - 136.85$ |

Notes: the model with fitness 5.98 is GP3. "Cost" is based on the number of summations. $f(r)$ is the smoothing function defined in the main manuscript

**Computational cost benchmarks of GP1 and GP2**

The interatomic potentials GP1 and GP2 were implemented in LAMMPS. The performance was measured in a system with 32 atoms for 10 million relaxation steps with a timestep of 1 fs in a single core. The paper by Sutton and Chen does not report a cutoff radius, but we found that a radius of at least 10 Å was required to reproduce their results. If we use the same 5 Å radius as used for GP1 and GP2, Sutton Chen EAM and GP1 have similar speeds because both are EAM-type potentials with 2 summations. EAM1 is slower than GP1 (Figure S1) because it used a greater cutoff distance of 5.50679 Å. GP2 is slower because it has 3 summations. However the difference in speed between the potentials compared in Figure S1 is small compared to the difference in speed between EAM and other potential models, which can be several orders of magnitude.[23,24]. The computational cost was measured on a single core of a Haswell node with a clock speed of 2.5 GHz.

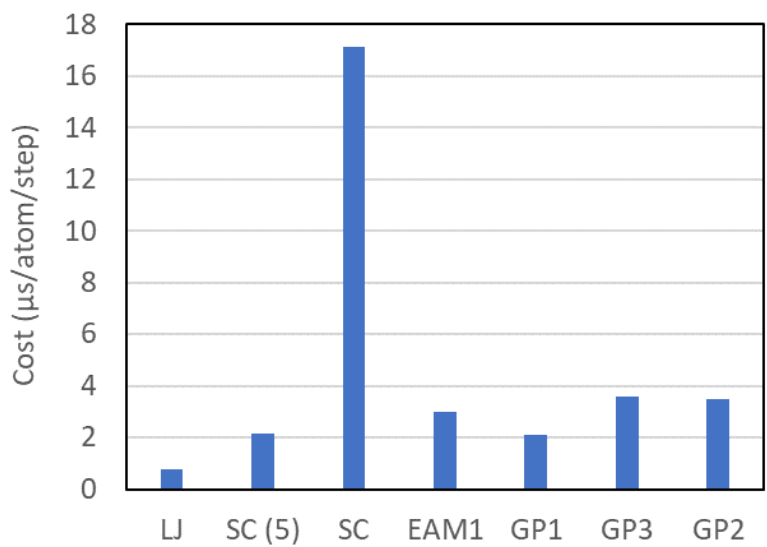

*Figure S1*. Computational cost of potential models in LAMMPS. SC (5) uses a cutoff distance of 5 Å, SC uses a cutoff distance of 10  . The cost is similar for SC (5), EAM1, GP1, GP2 and GP3.

**Mean absolute percent error on elastic constants**

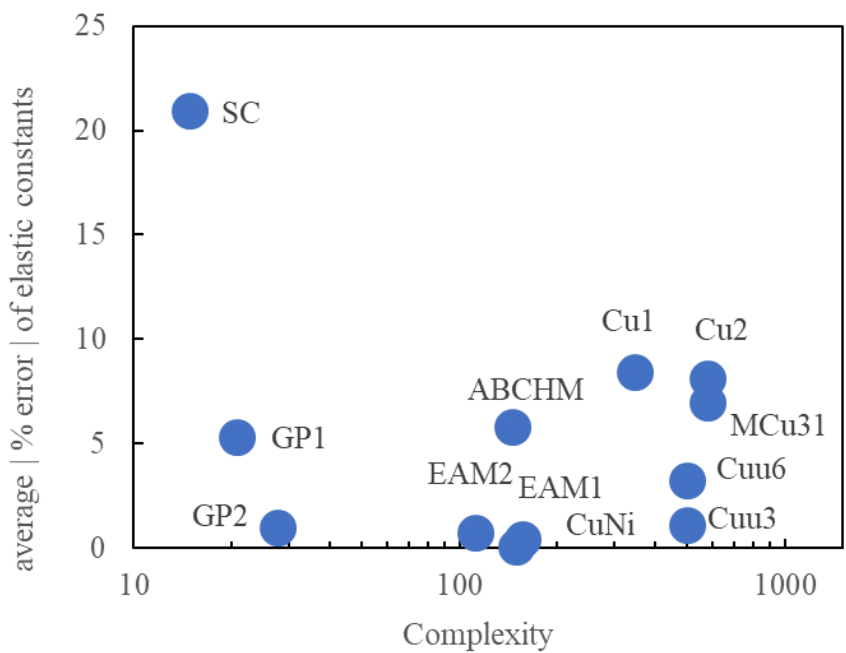

*Figure S2.* Average (instead of maximum) absolute error on elastic constants. The Pareto frontier does not change from the one calculated using maximum error.

**Enabling GP1, GP2 and GP3 in LAMMPS**

GP1 can be used under the "pair_style eam/alloy" in LAMMPS. Cu_GP1.eam is the corresponding potential file. The following is an example input specification:

    pair_style eam/alloy

    pair_coeff * * Cu_GP1.eam Cu

GP2 and GP3 use "pair_style poet" and require the potential file Cu_GP2.poet or Cu_GP3.poet, respectively. The following is an example input specification:

    pair_style poet

    pair_coeff * * Cu_GP2.poet Cu

The poet pair_style can be compiled in LAMMPS following these steps:

1. Copy the files "pair_poet.cpp" and "pair_poet.h" (available in the Supplementary Information) to <lammps_main_directory>/src/MANYBODY and edit the file <lammps_main_directory>/src/Makefile.list by adding "pair_poet.cpp" and "pair_poet.h" to the end of the respective lines
2. *make* LAMMPS by including the "yes-manybody" flag